\documentclass[aps,prb,twocolumn,amsfonts,superscriptaddress]{revtex4-1}

\usepackage[latin1]{inputenc}
\usepackage{amsmath,amssymb}
\usepackage{graphicx,color}
\usepackage{braket}
\usepackage{subfigure}
\usepackage{bm}
\usepackage{mathtools}

\usepackage{titlesec}

\usepackage[breaklinks]{hyperref}
\usepackage[dvipsnames]{xcolor}
\hypersetup{colorlinks,citecolor=blue,filecolor=blue,linkcolor=blue,urlcolor=blue}

\usepackage{natbib}
\setcitestyle{square,numbers}

\newcommand{\hc}[0]{\textrm{H.c.}}
\newcommand{\mf}[1]{\mathfrak{#1}}
\newcommand{\mc}[1]{\mathcal{#1}}
\newcommand{\mb}[1]{\mathbb{#1}}
\newcommand{\trm}[1]{\textrm{#1}}

\newcommand{\tbf}[1]{\textbf{#1}}
\renewcommand{\vec}[1]{\bm{#1}}

\allowdisplaybreaks[0]

\begin{document}

\title{Hinge states in a system of coupled Rashba layers}

\author{Kirill Plekhanov, Flavio Ronetti, Daniel Loss, and Jelena Klinovaja}
\affiliation{Department of Physics, University of Basel,
  Klingelbergstrasse 82, CH-4056 Basel, Switzerland}
\date{\today}

\begin{abstract}
  We consider a system of stacked tunnel-coupled two-dimensional
  electron- and hole-gas layers with Rashba spin-orbit interactions
  subjected to a staggered Zeeman field. The interplay of different
  intra-layer tunnel couplings results in a phase transition to a
  topological insulator phase in three dimensions hosting gapless
  surface states. The staggered Zeeman field further enriches the
  topological phase diagram by generating a second-order topological
  insulator phase hosting gapless hinge states. The emergence of the
  topological phases is proven analytically in the regime of small
  Zeeman field and confirmed by numerical simulations in the
  non-perturbative region of the phase diagram. The topological phases
  are stable against  external perturbations and  disorder.
\end{abstract}

\maketitle

\section{Introduction \label{sec:introduction}}

Over the last decade, topological insulators and superconductors (TIs)
have attracted a lot of attention in the domain of condensed matter
physics~\cite{TIRev_HasanKane2010, TIRev_QiZhang2011,
  TIRev_SatoAndo2017, TIRev_WangZhang2017, TIRev_Wen2017}. One of the
main reasons for such popularity resides in the bulk-boundary
correspondence relating the topological description of a
$d$-dimensional insulating bulk to the presence of gapless modes on a
$(d-1)$-dimensional boundary of TIs. Moreover, the topological nature
of the boundary modes makes them insensitive to a wide spectrum of
external perturbations and disorder -- the property of greatest
importance in the area of quantum information and quantum
computing~\cite{QComp_FreedmanKitaevLarsenWang2003, QComp_Kitaev2003,
  QComp_SternLindner2013}. Recently, the concept of bulk-boundary
correspondence has been generalized by introducing a new class of
topological systems, called higher-order
TIs~\cite{HOTI_BenalcazarBernevigHughes2017,
  HOTI_BenalcazarBernevigHughes2017_2, HOTI_SongFangFang2017}. In
contrast to conventional TIs, the $(d-1)$-dimensional boundary of an
$n$-th order TI is insulating, similarly to the bulk. Instead, it
exhibits protected gapless modes on the $(d-n)$-dimensional
boundaries. The corresponding gapless modes are called corner states
in the case $d\geq2$, $n=d$ or hinge states for $d\geq3$, $n=d-1$.

Following the pioneering works on higher-order TIs, a great success
has been achieved in understanding these states of
matter~\cite{PengBaoOppen2018, GeierEtAl2018,
 HsuStanoKlinovajaLoss2018, Liu2018,
 VolpezLossKlinovaja2018, Ezawa2018b, roy1, roy2, kat, kirill,refTI,refTI2,refTI3,refTI4,refTI5}. In two
dimensions, the second-order TIs hosting corner states have been
realized experimentally using electromagnetic
circuits~\cite{PetersonBenalcazarHughesBahl2018,
  ImhofEtAl2018}, photonic lattices~\cite{MittalEtAl2019,
  ChenEtAl2019} and waveguides~\cite{HassanKunstEtAl2019}, as well as
phononic~\cite{SerraGarciaEtAl2018} and
acoustic~\cite{XueYangEtAl2019} setups. In three dimensions, there are
strong indications that certain materials, such as
$\trm{SnTe}$~\cite{SchindlerEtAl2018_1} and
bismuth~\cite{SchindlerEtAl2018_2}, behave as second-order TIs hosting
hinge states. However, only a few experimental realizations of
higher-order TIs with a high level of control over the system
parameters exist in $d\geq3$~\cite{WeinerNiLiAluKhanikaev2019,
  ZhangXieEtAl2019}.

\begin{figure}[t]
  \centering
  \includegraphics[width=.75\columnwidth]
  {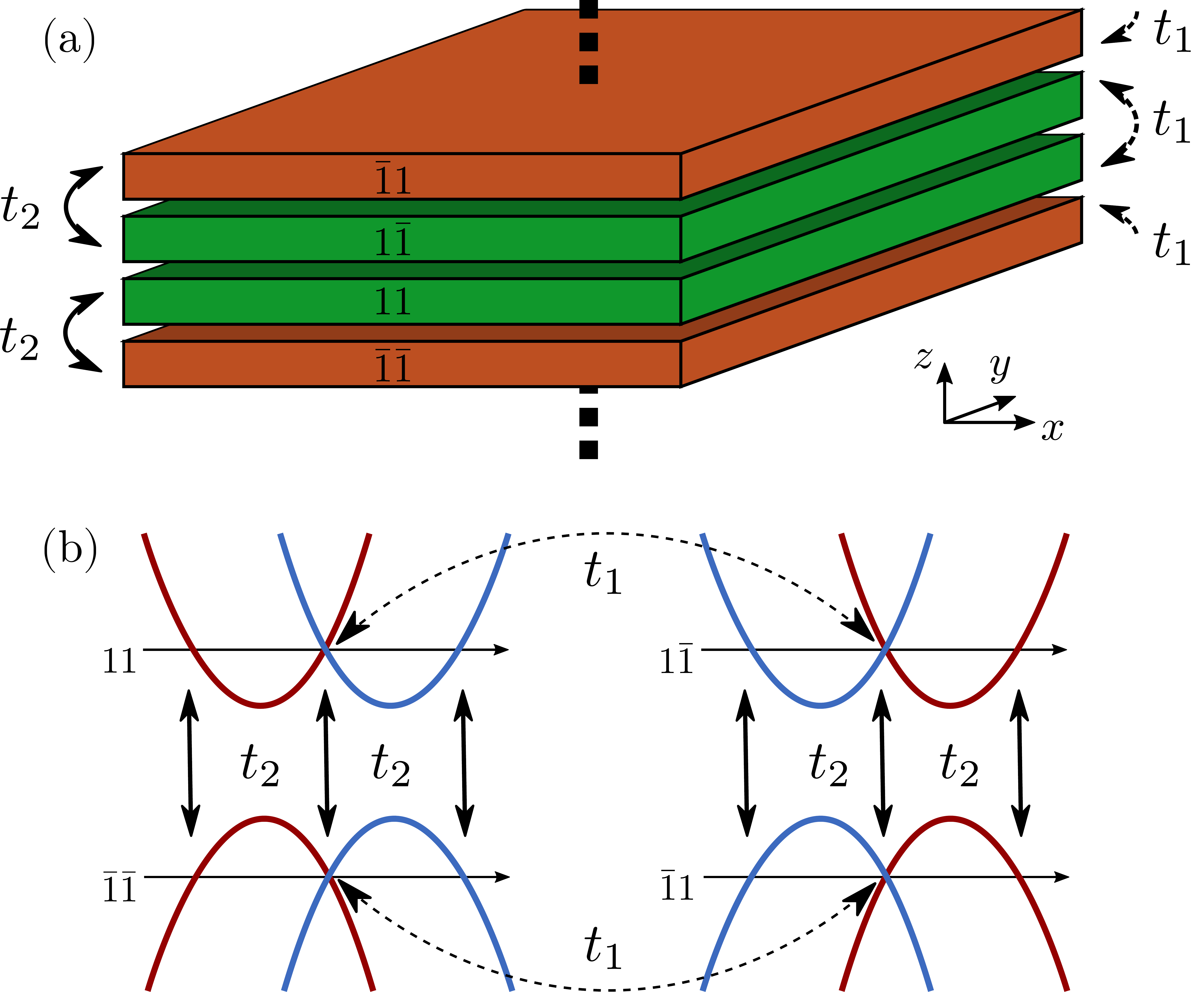}
  \caption{(a) Schematic representation of the system of coupled
    Rashba layers. Different layers are described by different
    dispersion types, indexed by
    $\eta \in \lbrace 1, \bar{1} \rbrace$, and different signs of the
    SOI amplitude, indexed by $\tau \in \lbrace 1, \bar{1}
    \rbrace$. The layers of same (different) dispersion type (represented by
    the same color) are coupled via a $t_1$ ($t_2$) tunneling term. 
    %The layers
    %of different dispersion types are coupled via $t_2$ tunneling term. 
    (b) A scheme showing different types of Rashba dispersions
    along the line $k_y = 0$ in  momentum space. The tunneling
    process $t_1$ ($t_2$) coupling different Rashba layers is
    represented by dashed (solid) lines.}
  \label{fig:model}
\end{figure}

Motivated by this context,  we propose here a mesoscopic setup
that can be controllably brought into the second-order TI (SOTI) phase
in three dimensions. The key component of our setup is an array of
two-dimensional electron gas (2DEG) layers with Rashba spin-orbit
interaction (SOI) and electron/hole-like
dispersions~\cite{Trifun0vicLossKlinovaja2016,
  VolpezLossKlinovaja2017} (see Fig.~\ref{fig:model}). Different
Rashba layers are tunnel coupled to each other, and the whole
heterostructure is subjected to a staggered Zeeman field.  The phase
diagram of the resulting model comprises a strong three-dimensional
topological insulator (3DTI) phase at zero Zeeman field, which
transforms into the SOTI phase when the Zeeman field is switched
on. In this work we do not include the effect of interactions, but we
point out the possibility to create even more exotic fractional
higher-order topological states in a 
%{\color{red}strongly correlated ??} 
version of
our system that contains strong electron-electron interactions~\cite{Trifun0vicLossKlinovaja2016,VolpezLossKlinovaja2017}.
%which cannot be achieved in semi-classical realizations of the TIs. 

This work is organized as follows. First, in Sec.~\ref{sec:model}
we introduce the model for coupled Rashba layers. Second, in
Sec.~\ref{sec:3DTI} we briefly present the phase diagram of the
model in the regime of a vanishing staggered Zeeman field. We show
that the system exhibits a phase transition from the topologically
trivial phase to the 3DTI phase hosting gapless surface states. Next,
in Sec.~\ref{sec:SOTI} we analyze the effect of a staggered Zeeman
field. We show that it gaps out the surface states and leads to the
emergence of the SOTI phase hosting gapless hinge states. We also
calculate the phase diagram of the model and study the stability of
the topological phases against the disorder. Finally, in
  Sec.~\ref{sec:surfaceField} we propose two modifications of the
  setup, which could facilitate the experimental realization. In the
  first proposal, the staggered Zeeman field is generated only on the
  surface of the heterostructure, using an antiferromagnet with a
  bicollinear antiferromagnetic order or via magnetic impurities. In
  the second proposal, the pairs of the 2DEGs are replaced by thin TI
  films.

\section{Model \label{sec:model}}

We start by introducing a system of coupled Rashba 2DEG layers
\cite{Trifun0vicLossKlinovaja2016, VolpezLossKlinovaja2017} placed in
the $xy$ plane (see Fig.~\ref{fig:model}).  We assume that the
corresponding SOI vector is along the $z$ axis and defines the spin
quantization axis.  The unit cell is composed of two pairs of layers
with electron- and hole-like dispersions described by opposite signs of
the mass $m$ and labeled by an index
$\eta \in \left\lbrace 1, \bar{1} \right\rbrace$. The layers of the
same dispersion type are distinguished by the value of the SOI
strength, which have the same amplitude $\alpha$ but opposite signs
indexed by $\tau = \left\lbrace 1, \bar{1} \right\rbrace$. Creation
operators $\psi^{\dag}_{\eta \tau \sigma \vec{k}}$ act on electrons
with spin component $\sigma$ along the $z$ axis and momentum
$\vec{k} = (k_x, k_y, k_z) \equiv (\vec{k}_{\parallel}, k_z)$ in the
layer $(\eta, \tau)$. Using this notation, the Hamiltonian describing
the uncoupled layers reads
\begin{align}
  &H_0
  =
    \sum_{\eta \tau \sigma \sigma'}
    \int \trm{d}^3 \vec{k}\
    \psi^{\dag}_{\eta \tau \sigma \vec{k}}
    \left[ \mc{H}^0_{\eta \tau}(\vec{k}) \right]_{\sigma \sigma'}
    \psi_{\eta \tau \sigma' \vec{k}}
    ,
    \notag \\
  &\mc{H}^{0}_{\eta \tau}(\vec{k})
  = 
    \left(
    \frac{\hbar^2 \vec{k}_{\parallel}^2}{2 m} - \mu
    \right) \eta +
    \alpha \tau \left( \sigma_x k_y - \sigma_y k_x \right)
    ,
\end{align}
where the Pauli matrices $\sigma_j$ act in  spin space. The chemical
potential $\mu$ is tuned to the crossing point of the spin-split bands
at $\vec{k}_{\parallel} = 0$. In the following, it will  be
convenient to introduce the SOI energy
$E_{\trm{so}} = \hbar^2 k_{\trm{so}}^2 / (2m)$ and the SOI momentum
$k_{\trm{so}} = m \alpha / \hbar^2$.

The nearest-neighbor Rashba layers are coupled to each other via
spin-conserving tunneling processes. We distinguish two different
types of tunnelings: between the layers with the same mass with the
amplitude $t_1$ and between the layers with opposite masses with the
amplitude $t_2$. The corresponding processes are described by the
following terms in the Hamiltonian 
\begin{align}
  & H_1
    = t_1 \sum\limits_{\sigma} \int \trm{d}^3 \vec{k}\
    \Big(
    \psi^{\dag}_{1 1 \sigma \vec{k}} \psi_{1 \bar{1} \sigma \vec{k}}
    \notag \\
  & \hspace{80pt} + e^{i a k_z}
    \psi^{\dag}_{\bar{1} 1 \sigma \vec{k}} \psi_{\bar{1} \bar{1} \sigma \vec{k}} +
    \hc
    \Big), \\
  & H_2
    = t_2 \sum\limits_{\sigma} \int \trm{d}^3 \vec{k}\
    \Big(
    \psi^{\dag}_{1 1 \sigma \vec{k}} \psi_{\bar{1} \bar{1} \sigma \vec{k}} +
    \psi^{\dag}_{1 \bar{1} \sigma \vec{k}} \psi_{\bar{1} 1 \sigma \vec{k}} +
    \hc
    \Big) \notag ,
\end{align}
where $a$ denotes the size of the unit cell in $z$ direction.

Finally, the system is subjected to an effective Zeeman term
\begin{align}
  H_{\trm{Z}}
  = t_{\trm{Z}} \sum\limits_{\eta \tau \sigma \sigma'}
  \eta 
  \int \trm{d}^3 \vec{k}\
  \psi^{\dag}_{\eta \tau \sigma \vec{k}}
  \left[ \vec{n} \cdot \vec{\sigma}
  \right]_{\sigma \sigma'}
  \psi_{\eta \tau \sigma' \vec{k}} .
\end{align}
Here, $t_{\trm{Z}} \eta n_j = \mu_{\trm{B}} g_j B_j$, with $\vec{B}$ --
the magnetic field vector, $\vec{n}$ -- the direction of the Zeeman
field, and $g_j$ -- the $g$-factor along a given axis.  We assume that the
sign of the Zeeman splitting alternates between the Rashba layers with
electron- and hole-like dispersions. Such a staggered effective Zeeman
field can be achieved in several ways. The first option is to use a
uniform magnetic field and  2DEGs with opposite signs of the
$g$-factor in layers with different dispersions. A second possibility
consists of using magnetic impurities which order ferromagnetically
within the same Rashba layers but have a staggered magnetization
direction in different layers~\cite{Imps_LiuEtAl2009,
  Imps_ChenEtAl2010, Imps_BurkovBalents2011,chen,takis}. Yet another option is to
use thin ferromagnetically ordered layers, contained between the 2DEGs
with the same dispersions. Alternatively, the effective Zeeman field
can be applied only to the surface of the system, for example by
placing it in the vicinity of a magnetic material, or by doping the
surface with magnetic adatoms. As we will show later in
Sec.~\ref{sec:surfaceField}, these constructions allow us to reach the desired
topological phases without affecting the bulk of the system.

After combining all previously introduced ingredients, the total
Hamiltonian can be conveniently written in the basis $\Psi_{\vec{k}}$
= $(\psi_{1 1 \uparrow \vec{k}}$, $\psi_{1 1 \downarrow \vec{k}}$,
$\psi_{1 \bar{1} \uparrow \vec{k}}$,
$\psi_{1 \bar{1} \downarrow \vec{k}}$,
$\psi_{\bar{1} 1 \uparrow \vec{k}}$,
$\psi_{\bar{1} 1 \downarrow \vec{k}}$,
$\psi_{\bar{1} \bar{1} \uparrow \vec{k}}$,
$\psi_{\bar{1} \bar{1} \downarrow \vec{k}})^T$ as
\begin{align}
  \label{eq:totalHamiltonian}
  H
  & =
    \int \trm{d}^3 \vec{k}\
    \Psi^{\dag}_{\vec{k}} \left(
    \mc{H}_{0} + \mc{H}_{1} + \mc{H}_{2} + \mc{H}_{\trm{Z}}
    \right) \Psi_{\vec{k}}
    \notag , \\
  \mc{H}_0
  & =
    \frac{\hbar^2 \vec{k}_{\parallel}^2}{2 m}
    \eta_z +
    \alpha \tau_z \left( \sigma_x k_y - \sigma_y k_x \right)
    , \notag \\
  \mc{H}_1
  & =
    t_1 \left(
    \frac{\eta_0 + \eta_z}{2} \tau_x +
    \frac{\eta_0 - \eta_z}{2}
    \left[ \cos \left( a k_z \right) \tau_x -
    \sin \left( a k_z \right) \tau_y
    \right]
    \right)
    , \notag \\
  \mc{H}_2
  & =
    t_2 \eta_x \tau_x
    , \quad
    \mc{H}_{\trm{Z}}
    = 
    t_{\trm{Z}}\, \eta_z\, \vec{n} \cdot \vec{\sigma},
\end{align}
where the Pauli matrices $\eta_j$ and $\tau_j$ act in the space of four layers
$(\eta, \tau)$. In the following, we will present the solution of the
problem described by the Hamiltonian $H$.

% In the following we will show a solution to this
%problem.

\section{Strong 3D topological insulator \label{sec:3DTI}}

In order to describe the properties of our model, we first focus on
the regime of vanishing Zeeman field ($t_{\trm{Z}} = 0$). In this
section we calculate the topological phase diagram and deduce the
effective degrees of freedom emerging at low energies, associated with
the topological surface states.

\subsection{Topological phase diagram}

When the Zeeman field is absent, the system is described by the
Hamiltonian density
\begin{align}
  \mc{H}_{\trm{TI}} = \mc{H}_0 + \mc{H}_1 + \mc{H}_2,
\end{align}
with the bulk spectrum given by
\begin{align}
  E^2 (\vec{k})
  & =
    \epsilon^2 + (\alpha \vec{k}_{\parallel})^2 + t_1^2 + t_2^2
    \notag \\
  & \pm
    2 \sqrt{\epsilon^2 (\alpha \vec{k}_{\parallel})^2 + \epsilon^2 t^2_1 +
    t_1^2 t_2^2 \cos^2 (a k_z / 2)} ,
\end{align}
where $\epsilon = \hbar \vec{k}_{\parallel}^2 / (2 m)$. The spectrum
does not depend on the direction of the vector $\vec{k}_{\parallel}$
as a consequence of the rotational symmetry in the $xy$ plane. We also
see that at zero momentum, $\vec{k} = 0$, the two tunneling processes
compete with each other, leading to the closing of the gap at the
critical point $t_1 = t_2$. As a result, the phase diagram is divided
into two regimes: $t_1 < t_2$ and $t_1 > t_2$.

\begin{figure}[t]
  \centering
  \includegraphics[width=.99\columnwidth]
  {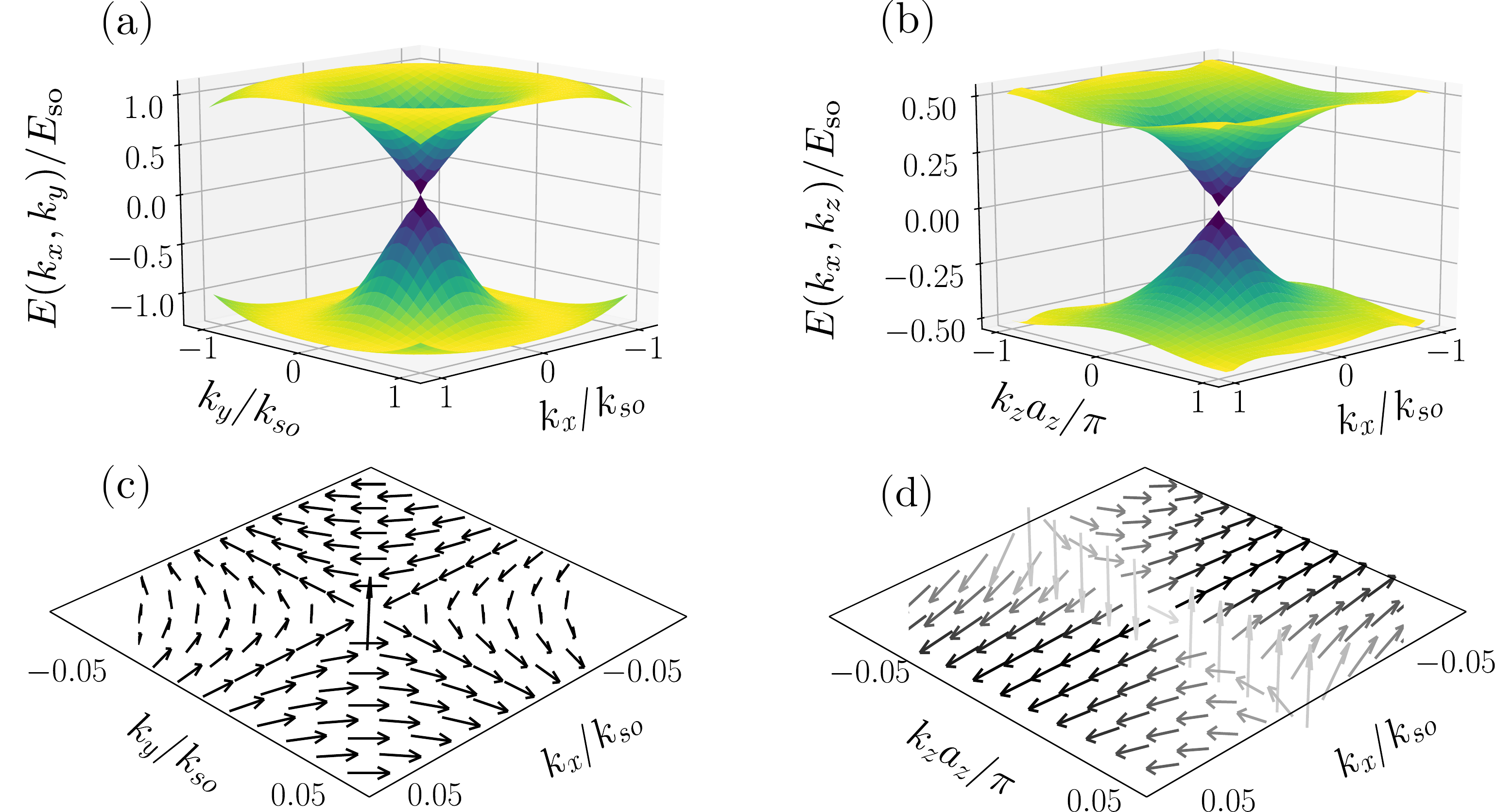}
  \caption{Dirac cone dispersion relation of the surface states (a)
    $\Ket{\Phi^{xy}_{\pm}}$ obtained in numerical simulations with PBC
    along $x$ and $y$ and OBC along $z$, and (b)
    $\Ket{\Phi^{xz}_{\pm}}$ obtained in numerical simulations with PBC
    along $x$ and $z$ and OBC along $y$.
%    , shown in units of
%    $E_{\trm{so}}$. 
    Both Dirac cones are centered at zero
    momentum. Panel (c) shows the spin-orbit polarization in
    momentum space of the states 
    $\Ket{\Phi^{xy}_{+}}$ in the surface layer at $z=0$ and panel (d) those of
    $\Ket{\Phi^{xz}_{+}}$ in the surface layer at $y=0$. The arrows represent the direction, while the
    color intensity represents the absolute value of the spin-orbit
    polarization. Parameters of the simulations are
    $t_1 = 2 t_2 = E_{\trm{so}}$.}
  \label{fig:surfaceStates_gapless_ens}
\end{figure}

The Hamiltonian density satisfies the relation
$T \mc{H}_{\trm{TI}}(\vec{k}) T^{-1} = \mc{H}_{\trm{TI}}(-\vec{k})$,
implying that the system is time-reversal symmetric. Here the
time-reversal symmetry operator is expressed as $T = i \sigma_y K$,
with $K$ being the complex conjugation operator. However, we note that
the particle-hole symmetry and the chiral symmetry are absent. Hence,
the system belongs to the $\trm{AII}$ symmetry class characterized by
a topological invariant $\nu_0 \in \mb{Z}_2$. The bulk-boundary
correspondence relates this topological invariant to the presence of
gapless states $\Ket{\Phi^{s}_{\pm}}$ localized at every surface $s$
of the system and decaying exponentially into the bulk. We verify
numerically the presence of such gapless surface states in the regime
$t_1 < t_2$ of our model by diagonalizing the discretized version of
the Hamiltonian $\mc{H}_{\trm{TI}}$ density, as shown in
Figs.~\ref{fig:surfaceStates_gapless_ens}(a)-(b). This allows us to
identify the regime $t_1 < t_2$ ($t_1 > t_2$) with the topologically
trivial (topological) phase, where the topological phase corresponds
to a strong 3DTI~\cite{Trifun0vicLossKlinovaja2016,
  VolpezLossKlinovaja2017}.

We study the surface states more closely by considering several
non-equivalent geometries of the system with open boundary conditions
(OBCs) along one direction and periodic boundary conditions (PBCs)
along two other directions. In particular, we study the $xy$ surface
states in a geometry with OBC along $z$ and PBCs along $x$ and $y$
[see Fig.~\ref{fig:surfaceStates_gapless_ens}(a)], and the $xz$
surface in a geometry with OBC along $y$ and PBCs along $x$ and $z$
[see Fig.~\ref{fig:surfaceStates_gapless_ens}(b)]. We observe that the
surface states have a linear dispersion relation with one Dirac cone
per surface. The shape of the Dirac cone and the spin texture of the
surface states change depending on the surface considered. As a
consequence of the rotational symmetry in the $xy$ plane, the Dirac
cone on the $xy$ surface is isotropic, and the spin texture is
anti-vortex like [see Fig.~\ref{fig:surfaceStates_gapless_ens}(c)].
With our choice of the spin quantization axis being the $z$ axis, the
spin-orbit polarization of the surface states $\Ket{\Phi^{xy}_{\pm}}$
lie in the $xy$ plane. In contrast, the Dirac cone on the $xz$ surface
is tilted. Moreover, the spin texture of the states
$\Ket{\Phi^{xz}_{\pm}}$ is dominated by the $x$ component of the
spin-orbit polarization [see
Fig.~\ref{fig:surfaceStates_gapless_ens}(d)] almost everywhere except
on the line $k_x = 0$, where the polarization along the $z$ axis becomes
dominant.

\subsection{Gapless surface states}

In order to acquire a better theoretical understanding of the 3DTI
phase, we calculate the wavefunctions of the states
$\Ket{\Phi^{s}_{\pm}}$ situated at different surfaces $s$. We solve
the Schroedinger equation obtained by making the substitution
$k_j \rightarrow -i \hbar \partial_j$ with
$j \in \lbrace x, y, z \rbrace$. For clarity, we only focus on the
solutions at the Dirac point, which are obtained by taking two of
three components of the momentum $\vec{k}$ to zero. The effect of the
neglected terms is treated perturbatively, based on the solutions at
the Dirac point.

\subsubsection{Top and bottom surfaces\label{sec:surfaceStatesTopZero}}

First we focus on the $xy$ surface by taking
$\vec{k}_{\parallel} = 0$. We rewrite the problem in a geometry with
OBC along the $z$ axis by substituting
$k_z \rightarrow - i \hbar \partial_z$. We assume that the linear size
$L_z$ of the system in $z$ direction is sufficiently large, such
  that the surface states at opposite surfaces can be treated as
  independent. The resulting simplified real space Hamiltonian
density reads
\begin{align}
  \label{eq:surfaceState_xy}
  \mc{H}^{xy}_{\trm{TI}}
  =
  t_2 \eta_x \tau_x + t_1 \tau_x
  + i t_1 \hbar a \partial_z  \frac{\eta_0 - \eta_z}{2} 
  \tau_y
  \;.
\end{align}
Here, we only consider the terms linear in momentum $k_z$. We
note that, in addition to a time-reversal symmetry $T$ already
present in the full problem, the simplified Hamiltonian has a
particle-hole symmetry represented by an operator $P =-\tau_z K$
satisfying
$P \mc{H}^{xy}_{\trm{TI}} P^{-1} = - \mc{H}^{xy}_{\trm{TI}}$ 
and
$\left[ P, T\right] = 0$. Moreover, the chiral symmetry described by
an operator $C = T P = - i \tau_z \sigma_y$ is also present in the
system. Enforced by these symmetries, the Dirac point of the surface
states is located exactly at zero energy.

Solving the Schroedinger equation for $\mc{H}^{xy}_{\trm{TI}}$ results
in two zero-energy solutions localized at the top surface at $z = 0$:
\begin{align}
  \Ket{\Phi^{xy}_{\pm}}
  &=
    \frac{1}{\mc{N}}
    \sum\limits_{ \eta \tau \sigma}
    %\int
    \int\limits_{z=0}^{L_z}
    \trm{d} z\
    \Phi^{xy}_{\pm \eta \tau \sigma}(z)
    \Ket{z, \eta, \tau, \sigma}
    , \notag \\
  \Phi^{xy}_{+}(z)
  &=
    e^{- z / \xi}
    (0, 0, 1, 0, 0, 0,-t_1/t_2, 0)
    , \notag \\
  \Phi^{xy}_{-}(z)
  &=
    e^{- z / \xi}
    (0, 0, 0, -1, 0, 0, 0, t_1/t_2) ,
\end{align}
where $\xi = t_1 t_2 \hbar a / (t_1^2 - t_2^2) $ and $\mc{N}$ is a
normalization constant. We remind that we work in the basis
  $\Phi^{xy}_{\pm}$ = $(\Phi^{xy}_{\pm 1 1 \uparrow}$,
  $\Phi^{xy}_{\pm 1 1 \downarrow}$,
  $\Phi^{xy}_{\pm 1 \bar{1} \uparrow}$,
  $\Phi^{xy}_{\pm 1 \bar{1} \downarrow}$,
  $\Phi^{xy}_{\pm \bar{1} 1 \uparrow}$,
  $\Phi^{xy}_{\pm \bar{1} 1 \downarrow}$,
  $\Phi^{xy}_{\pm \bar{1} \bar{1} \uparrow}$,
  $\Phi^{xy}_{\pm \bar{1} \bar{1} \downarrow})^T$. The solutions
satisfy the relations
$T \Ket{\Phi^{xy}_{\pm}} = \pm \Ket{\Phi^{xy}_{\mp}}$ and
$P \Ket{\Phi^{xy}_{\pm}} = \Ket{\Phi^{xy}_{\pm}}$. More generally, any
combination of $\Ket{\Phi^{xy}_{+}}$ and $\Ket{\Phi^{xy}_{-}}$ (with
its time-reversal partner) is also a solution of the problem, however,
only these two states respect the rotational symmetry in the $xy$
plane present in $\mc{H}_{\trm{TI}}$.

The dispersion relation and the effective Hamiltonian at finite
momentum $\vec{k}_{\parallel}$ is obtained perturbatively by
calculating the expectation values of the remaining terms in
$\mc{H}_{\trm{TI}}$ in the basis of the surface states
$\Ket{\Phi^{xy}_{\pm}}$. Here we only describe the low-energy physics
close to the Dirac point, assuming that the amplitude of the momentum
$\vec{k}_{\parallel}$ is small. Hence, we only take into account the
terms linear in $k_x$ and $k_y$. We also distinguish between the
diagonal and off-diagonal components of the expectation values. We
find that all the diagonal components vanish exactly, i.e.
\begin{align}
  &
    \Braket{\Phi^{xy}_{\pm} |
    \alpha k_y \sigma_x \tau_z |
    \Phi^{xy}_{\pm}} =
    \Braket{\Phi^{xy}_{\pm} |
    \alpha k_x \sigma_y \tau_z |
    \Phi^{xy}_{\pm}} = 0 .
\end{align}
However, the off-diagonal components are non-zero:
\begin{align}
  & \Braket{\Phi^{xy}_{-} |
    \alpha k_y \sigma_x \tau_z |
    \Phi^{xy}_{+}}
    = \alpha k_y (t_1 - t_2) / \bar{t},
    \notag \\
  & \Braket{\Phi^{xy}_{-} |
    \alpha k_x \sigma_y \tau_z |
    \Phi^{xy}_{+}}
    = i \alpha k_x (t_1 - t_2) / \bar{t},
    \notag \\
  & \bar{t} =
    2 \mc{N} \sqrt{t_2^2 + t_1^2}
    \left[ {\xi \left( 1 -  e^{{- 2 L_z / \xi}} \right)} \right]^{-1}
    .
\end{align}
This allows us to write the effective low-energy Hamiltonian density
of the top $xy$ surface as
\begin{align}
  \mc{H}^{xy}_{\trm{TI}} (z=0)
  \approx \alpha \left( \rho_x k_y - \rho_y k_x \right)
  (t_1 - t_2) / \bar{t}
  ,
\end{align}
where $\rho_j$ act in the space of the surface states
$\Ket{\Phi^{xy}_{\pm}}$. We recover the Dirac cone dispersion relation
of the surface states and note that at finite momentum
$\vec{k}_{\parallel} \neq 0$ the spin-orbit polarization is
momentum-locked. Combined with the rotational symmetry in the $xy$
plane, this results in the anti-vortex like spin texture observed
numerically in Fig.~\ref{fig:surfaceStates_gapless_ens}(c).

The low-energy description of the bottom $xy$ surface at $z = L_z$ is
obtained using the symmetry transformation $z \rightarrow L_z - z$,
$\vec{e}_j \rightarrow -\vec{e}_j$, where $\vec{e}_j$ are three basis
unit vectors. Under this transformation, the Hamiltonian density is
modified as
$\mc{H}^{xy}_{\trm{TI}} \rightarrow \tau_x \mc{H}^{xy}_{\trm{TI}}
\tau_x$. Such a transformation preserves the orientation of the
surface, so that the outward-pointing normal vector
$\vec{v}^{xy}_{\perp}$ always coincides with the vector
$\vec{e}_z$. Using this symmetry transformation, we deduce that the
wavefunctions of the states localized at $z = L_z$ read
$\tau_x \Phi^{xy}_{\pm}(L_z-z)$. The effective low-energy Hamiltonian
expressed in the basis of the surfaces states remains unaffected,
\begin{align}
  \mc{H}^{xy}_{\trm{TI}} (z=L_z) =
  \mc{H}^{xy}_{\trm{TI}} (z=0).
\end{align}

\subsubsection{Lateral surfaces\label{sec:surfaceStatesLatZero}}

In the same way as for the top and bottom surfaces, we express the
states $\Ket{\Phi^{xz}_{\pm}}$ localized on the $xz$ surface in the
neighborhood of the Dirac point, by considering $(k_x, k_z) = 0$ and
making the substitution $k_y \rightarrow - i \hbar \partial_y$. We
  also assume that the linear size $L_y$ of the system in $y$
  direction is sufficiently large, such that the overlap between the
  states at opposite surfaces can be neglected. The corresponding
problem reads
\begin{align}
  \label{eq:surfaceState_xz}
  \mc{H}^{xz}_{\trm{TI}}
  =
  -\frac{\hbar^2}{2 m} \partial_y^2 \eta_z -
  i \hbar \alpha \partial_y \tau_z \sigma_x +
  t_2 \eta_x \tau_x +
  t_1 \tau_x
  \;.
\end{align}
We note that the simplified Hamiltonian density has a new
particle-hole symmetry $P = \eta_x \tau_z K$ and a new chiral symmetry
$C = i \eta_x \tau_z \sigma_y$, which both commute with $T$. As a
consequence, the lateral surface states also have to be at zero energy
and satisfy $T \Ket{\Phi^{xz}_{\pm}} = \pm \Ket{\Phi^{xz}_{\mp}}$ and
$P \Ket{\Phi^{xz}_{\pm}} = \Ket{\Phi^{xz}_{\pm}}$.

The zero-energy solutions of Eq.~\eqref{eq:surfaceState_xz} describing the
state localized at the $y=0$ surface have the following general form:
\begin{align}
  \Ket{\Phi^{xz}_{\pm}}
  =&
     \frac{1}{\mc{N}}
     \sum\limits_{\sigma \eta \tau} \int_{y}
     \trm{d} y\
     \Phi^{xz}_{\pm \eta \tau \sigma}(y)
     \Ket{y, \eta, \tau, \sigma}
     , \\
  \Phi^{xz}_{+}(y)
  =&
     \sum_{j}
     C_j
     e^{- y / \xi_j} e^{ i q_{j} y}
     \Phi^{xz}_{+j}
     , \notag \\
  \Phi^{xz}_{-}(y)
  =&
     \sum_{j}
     C^{*}_j
     e^{- y / \xi_j} e^{-i q_{j} y}
     \Phi^{xz}_{-j}
     , \notag
\end{align}
where $C_{j}$ are complex coefficients and $\Phi^{xz}_{\pm j}$ are
wavefunctions satisfying $T \Phi^{xz}_{\pm j} = \pm \Phi^{xz}_{\mp j}$,
$P \Phi^{xz}_{\pm j} = \Phi^{xz}_{\pm j}$, and
$\sigma_x \Phi^{xz}_{\mu j} = \pm \Phi^{xz}_{\mu j}$. Each
wavefunction $\Phi^{xz}_{\pm j}$ is associated with a decay length
$\xi_j$ and a wave-vector $q_j$, which can be calculated as
$\xi_j = 1 / \mf{Re}(\lambda_j)$ and $q_j = \mf{Im}(\lambda_j)$ \cite{composite}, where
$\lambda_j$ are solutions of
\begin{widetext}
  \begin{align}
    \label{eq:lambda_xz}
    \left( \frac{\hbar^2}{2m} \right)^4 \lambda^8 +
    2 \left( \frac{\hbar^2}{2m} \right) \left( \hbar \alpha \right)^2
    \lambda^6 +
    \left[
    \left( \hbar \alpha \right)^4 +
    \left( \frac{\hbar^2}{2m} \right)^2 (t_1^2 + t_2^2)
    \right]
    \lambda^4 -
    2 \left( \hbar \alpha \right)^2 (t_1^2 + t_2^2) \lambda^2
    + (t_1^2 - t_2^2)^2 - 4 t_1^2 \left( \frac{\hbar^2}{2 m} \right)^2 = 0 \;.
  \end{align}
\end{widetext}
Solutions of the above equation can be obtained analytically. For the
sake of readability, we do not provide an exact expression here,
however, we point out an important difference with respect to the
lateral surface case. The solutions of Eq.~\eqref{eq:lambda_xz} are,
in general, complex, such that both $\xi_j$ and $q_j$ are
non-zero. This corresponds to an exponentially decaying solution which
oscillates, as compared to the surface states on the top and bottom
surfaces which are simply decaying.

Due to the high complexity of the problem, we do not provide an
analytical expression of the wavefunctions $\Phi^{xz}_{\pm j}$ and the
coefficients $C_j$. Nevertheless, we are able to solve the
one-dimensional problem associated with Eq.~\eqref{eq:surfaceState_xz}
numerically. This allows us to estimate the expectation values of the
remaining terms in $\mc{H}_{\trm{TI}}$ in the basis of
$\Ket{\Phi^{xz}_{\pm}}$. In general, we find that the diagonal
components, as well as the real and the imaginary part of the
off-diagonal components, are non-zero for all the terms, as a result
of the strong anisotropy between the $x$ and $z$ directions. However,
we find that one of the contributions is usually by an order of
magnitude larger than the other terms. Below, we express the diagonal
components of the expectation values that accounts for this largest
contribution,
\begin{align}
  &
    \Braket{\Phi^{xz}_{\pm} |
    \alpha k_x \sigma_y \tau_z |
    \Phi^{xy}_{\pm}} = \pm \alpha A k_x, \quad A \in \mb{R}
    \notag \\
  &
    \Braket{\Phi^{xz}_{\pm} |
    t_1 a k_z (\eta_0 - \eta_z) \tau_y |
    \Phi^{xy}_{\pm}} = 0.
\end{align}
Here, $A$ is a coefficient that depends on the parameters of the
system and the constants $C_j$. Similarly, we calculate the
off-diagonal components
\begin{align}
  &
    \Braket{\Phi^{xz}_{-} |
    \alpha k_x \sigma_y \tau_z |
    \Phi^{xy}_{+}} = 0,
    \notag \\
  &
    \Braket{\Phi^{xz}_{-} |
    t_1 a k_z (\eta_0 - \eta_z) \tau_y |
    \Phi^{xy}_{+}} = t_1 B a k_z,\quad B \in \mb{R}.
\end{align}
As a result, the effective low-energy Hamiltonian of the surface
states at $y=0$ can be written as
\begin{align}
  \mc{H}^{xz}_{\trm{TI}} (y=0)
  \approx \alpha A k_x \rho_z + t_1 B a k_z \rho_x,
\end{align}
where $\rho_j$ act in the space of the surface states
$\Ket{\Phi^{xz}_{\pm}}$. We recover the tilted Dirac cone dispersion
relation observed in Fig.~\ref{fig:surfaceStates_gapless_ens}(d).

The effective Hamiltonian of the lateral surface at $y = L_y$ is
obtained by using the transformation $y \rightarrow L_y - y$,
$\vec{e}_j \rightarrow -\vec{e}_j$ which preserves the surface
orientation such that the normal vector $\vec{v}^{xz}_{\perp}$
pointing outwards of the surface coincides with the vector
$\vec{e}_y$. Under this transformation, the Hamiltonian density
transforms as
$\mc{H}^{xz}_{\trm{TI}} \rightarrow \sigma_y \mc{H}^{xz}_{\trm{TI}}
\sigma_y$, which allows us to express the surface state wavefunctions
as $\sigma_y \Phi^{xz}_{\pm}(L_y - y)$. We find that the low-energy
effective Hamiltonian density remains invariant under this
transformation
\begin{align}
  \mc{H}^{xz}_{\trm{TI}} (y = L_y) =
  \mc{H}^{xz}_{\trm{TI}} (y = 0)
  \;.
\end{align}

Finally, the low-energy description of the two remaining $yz$
  surfaces can be obtained using the rotational symmetry in the $xy$
  plane described by the transformation
  $\lbrace L_x-x, y \rbrace \rightarrow \lbrace y, x \rbrace$. Using
  this symmetry allows us to express the wavefunctions of the surface
  states localized at the $yz$ surfaces at $x = 0$ as
  $i \sigma_z \Phi^{xz}_{\pm}(x)$. Similarly, we find that the
  wavefunctions of the surface states localized at $x = L_x$ are
  $\sigma_x \Phi^{xz}_{\pm}(L_x - x)$. Under this
  transformation, the low-energy Hamiltonian describing the two $yz$
  surfaces becomes
  \begin{align}
    \mc{H}^{yz}_{\trm{TI}}(x=0) = \mc{H}^{yz}_{\trm{TI}}(x=L_x)
    \approx -\alpha A k_y \rho_z + t_1 B a k_z \rho_x
    .
  \end{align}
  Here we see that, the momentum component $k_x$ is replaced by $k_y$,
  but the structure of the low-energy Hamiltonian remains invariant.

\section{Second-order 3D topological insulator \label{sec:SOTI}}

In this section, we describe the low-energy physics of our system under
applied staggered Zeeman field ($t_{\trm{Z}} \neq 0$). For this
purpose, we consider the full Hamiltonian density
\begin{align}
  \mc{H}_{\trm{SOTI}} = \mc{H}_0 + \mc{H}_1 + \mc{H}_2 + \mc{H}_{\trm{Z}}
  \;.
\end{align}
The crucial role of the Zeeman term $\mc{H}_{\trm{Z}}$ consists in
gapping out the surface states $\Ket{\Phi^{s}_{\pm}}$. As a result,
the system ceases to be a 3DTI. Nevertheless, as we will explain below
in this section, the topological description of the resulting system
becomes even richer, supporting the emergence of chiral hinge states
characteristic to three-dimensional SOTI. We also note that the
spatial oscillation of the staggered Zeeman field in
$\mc{H}_{\trm{Z}}$ is crucial to achieve our goal. More precisely, one
can show that a uniform Zeeman field fails to gap out all the
surface states, and does not lead to  hinge states
(see~\ref{sec:app-UniformZeeman} for more details).

\subsection{Small Zeeman field}

In a first step, we consider the regime of a small staggered
Zeeman field $t_{\trm{Z}} \ll t_1, t_2$. This allows us to use the
results of Sec.~\ref{sec:3DTI} and treat the Zeeman term
$\mc{H}_{\trm{Z}}$ perturbatively.

\begin{figure*}
  \centering
  \includegraphics[width=1.99\columnwidth]
  {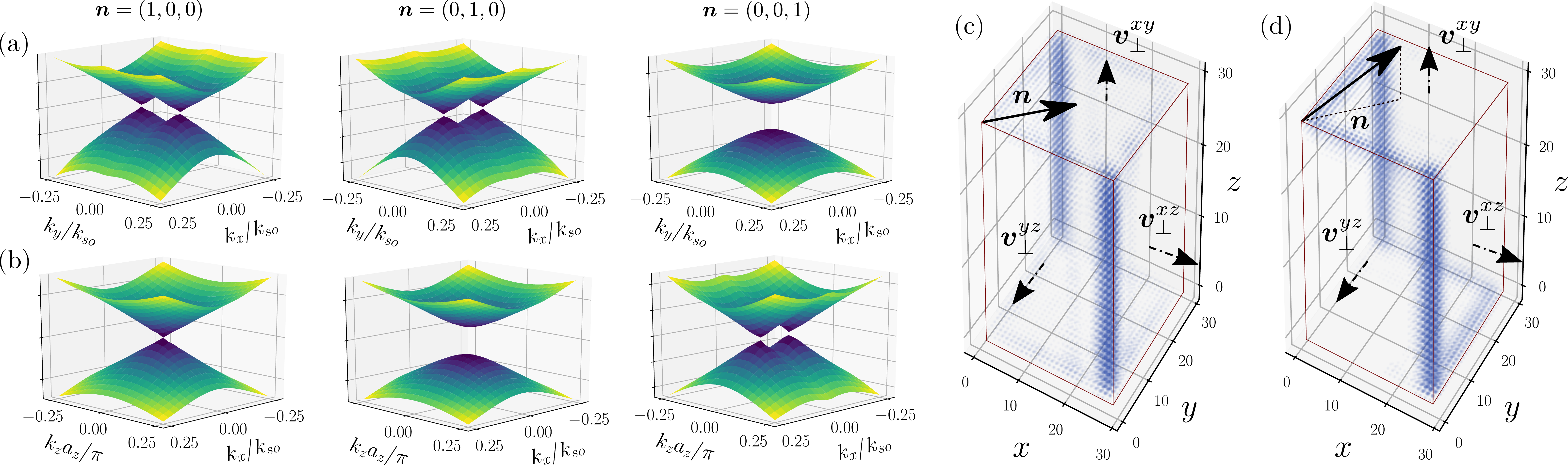}
  \caption{The effect of the staggered Zeeman field on the dispersion
    relation of the surface states (a) $\Ket{\Phi^{xy}_{\pm}}$ and (b)
    $\Ket{\Phi^{xz}_{\pm}}$. The direction of the Zeeman field is
    indicated by the vector ${\vec{n}}$. While the Zeeman field
    parallel to the surface either shifts the Dirac cone or does not affect
    the energy dispersion of the surface sates at all, the Zeeman
    field perpendicular to the surface opens a gap. Parameters of
    the simulations are $t_1 = 2 t_2 = E_{\trm{so}}$,
    $t_{\trm{Z}} = 0.3 E_{\trm{so}}$. (c)-(d) Probability density of
    the gapless hinge state obtained in numerical simulations with
    OBCs along $x$, $y$, and $z$. The sign of the mass term
      generated by the Zeeman field is equal to the sign of
      $\vec{v}^{s}_{\perp} \cdot \vec{n}$, where $\vec{v}^{s}_{\perp}$
      is the outward-pointing unit vector perpendicular to the surface
      $s$. (c) If $\vec{n} = (1, 1, 0) / \sqrt{2}$, the Zeeman field
    opens the gap at the surfaces $xz$ and $yz$ but leaves the surface
    $xy$ gapless. (d) If $\vec{n} = (1, 1, 1) / \sqrt{3}$, all the
    surfaces become gapped. In both cases hinge states emerge at the
    interface between gapped surfaces with opposite signs of the mass
    $M_s$. Parameters of the simulations are
    $t_1 = 2 t_2 = 4 E_{\trm{so}}$, $t_{\trm{Z}} = 2 E_{\trm{so}}$.}
  \label{fig:surfaceStates_gapped_ens}
\end{figure*}

\subsubsection{Top and bottom surfaces}

Using the expression of the wavefunctions $\Phi^{xy}_{\pm}(z)$ of the
surface states located on the top $xy$ surface at $z=0$, we calculate
the expectation values of the Zeeman term $\mc{H}_{\trm{Z}}$ in the
neighborhood of the Dirac point. We obtain the following diagonal
contributions
\begin{align}
  &
    \Braket{\Phi^{xy}_{\pm} |
    t_{\trm{Z}} n_x \sigma_x \eta_z |
    \Phi^{xy}_{\pm}} =
    \Braket{\Phi^{xy}_{\pm} |
    t_{\trm{Z}} n_y \sigma_y \eta_z |
    \Phi^{xy}_{\pm}} =
    0
    , \notag \\
  &
    \Braket{\Phi^{xy}_{\pm} |
    t_{\trm{Z}} n_z \sigma_z \eta_z |
    \Phi^{xy}_{\pm}} = \pm t_{\trm{Z}} n_z (t_1 + t_2) / \bar{t} .
\end{align}
Similarly, we calculate the off-diagonal contributions
\begin{align}
  &
    \Braket{\Phi^{xy}_{-} |
    t_{\trm{Z}} n_x \sigma_x \eta_z |
    \Phi^{xy}_{+}} = -
    t_{\trm{Z}} n_x (t_1 + t_2) / \bar{t}
    , \notag \\
  &
    \Braket{\Phi^{xy}_{-} |
    t_{\trm{Z}} n_y \sigma_y \eta_z |
    \Phi^{xy}_{+}} = - i t_{\trm{Z}} n_y (t_1 + t_2) / \bar{t}
    , \notag \\
  &
    \Braket{\Phi^{xy}_{-} |
    t_{\trm{Z}} n_z \sigma_z \eta_z |
    \Phi^{xy}_{+}} = 0
    .
\end{align}
Using these results, we write the effective low-energy Hamiltonian
density describing the top $xy$ surface of the 3DTI under the applied
Zeeman field as
\begin{align}
  \mc{H}^{xy}_{\trm{SOTI}}
  (z = 0)
  \approx
  \alpha
  \left( \rho_x k_y - \rho_y k_x \right)
  (t_1 - t_2) / \bar{t}
  &
    \notag \\
  -
  t_{\trm{Z}}
  \left( n_x \rho_x + n_y \rho_y - n_z \rho_z \right)
  (t_1 + t_2) / \bar{t}
  & .
\end{align}
We see that $x$ and $y$ components of the Zeeman field simply shift
the position of the Dirac cone in momentum space. However, $z$
component of the Zeeman field induces a 'mass term' with the amplitude
$M_{xy} = t_{\trm{Z}} n_z (t_1 + t_2) / \bar{t}$. This mass term
anticommutes with all the remaining terms in
$\mc{H}^{xy}_{\trm{SOTI}}$ and gaps out the surface states
$\Ket{\Phi^{xy}_{\pm}}$. These theoretical predictions are confirmed
by a numerical diagonalization of $\mc{H}_{\trm{SOTI}}$ presented in
Fig.~\ref{fig:surfaceStates_gapped_ens}(a).

Next, we perform the calculations for the bottom $xy$ surface at $z =
L_z$. We note that the projection of the Zeeman field vector
$\vec{n}$ onto the normal unit vector $\vec{v}^{xy}_{\perp}$, pointing
outwards of the surface, changes sign. In other words, applying the
surface-interchanging transformation from
Sec.~\ref{sec:surfaceStatesTopZero} results in the inversion of the
direction of the Zeeman field $\vec{n} \rightarrow - \vec{n}$. As a
consequence, the expectation value of the Zeeman term in the basis of
the surface states of the bottom surface is exactly opposite to the
case of the top $xy$ surface, leading to
\begin{align}
  \mc{H}^{xy}_{\trm{SOTI}}
  (z = L_z)
  \approx
  \alpha
  \left( \rho_x k_y - \rho_y k_x \right)
  (t_1 - t_2) / \bar{t}
  &
    \notag \\
  +
  t_{\trm{Z}}
  \left( n_x \rho_x + n_y \rho_y - n_z \rho_z \right)
  (t_1 + t_2) / \bar{t}
  & .
\end{align}
In the following, we will show an important consequence of this
result.

\subsubsection{Lateral surfaces}

Using the results of Sec.~\ref{sec:surfaceStatesLatZero}, we
analyze the effect of the Zeeman field on the surface states localized
on the lateral $xz$ surfaces at $y=0$. Below we provide the result of
a numerical calculation of the expectation value of the Zeeman term
$\mc{H}_{\trm{Z}}$ in the basis of states $\Ket{\Phi^{xz}_{\pm}}$.
For the diagonal components of the expectation values, we obtain
\begin{align}
  &
    \Braket{\Phi^{xz}_{\pm} |
    t_{\trm{Z}} n_x \sigma_x \eta_z |
    \Phi^{xz}_{\pm}} =
    \Braket{\Phi^{xz}_{\pm} |
    t_{\trm{Z}} n_y \sigma_y \eta_z |
    \Phi^{xz}_{\pm}} =
    0
    , \notag \\
  &
    \Braket{\Phi^{xy}_{\pm} |
    t_{\trm{Z}} n_z \sigma_z \eta_z |
    \Phi^{xy}_{\pm}} = \pm t_{\trm{Z}} n_z C, \quad C \in \mb{R}.
\end{align}
Similarly, for the off-diagonal components we find
\begin{align}
  &
    \Braket{\Phi^{xz}_{-} |
    t_{\trm{Z}} n_x \sigma_x \eta_z |
    \Phi^{xz}_{+}} =
    \Braket{\Phi^{xz}_{-} |
    t_{\trm{Z}} n_z \sigma_z \eta_z |
    \Phi^{xz}_{+}} =
    0
    , \notag \\
  &
    \Braket{\Phi^{xy}_{\pm} |
    t_{\trm{Z}} n_y \sigma_y \eta_z |
    \Phi^{xy}_{\pm}} = i t_{\trm{Z}} n_y C.
\end{align}
This allows us to write the effective low-energy Hamiltonian of the
$y=0$ surface as
\begin{align}
  \mc{H}^{xz}_{\trm{SOTI}}
  (y = 0)
  & \approx \alpha A k_x \rho_z + t_1 B a k_z \rho_x
    \notag \\
  & +
    t_{\trm{Z}} C
    \left( n_y \rho_y + n_z \rho_z \right) .
\end{align}
Surprisingly, we find that the Zeeman field along the $x$ direction
does not have any effect on the surface states. The $z$ component of
the Zeeman field simply shifts the Dirac cone along the $x$ axis in
the momentum space, while the $y$ component gaps out the surface
states by inducing a mass term with the amplitude
$M_{xz} = t_{\trm{Z}} C n_y$. We confirm these analytical
predictions by performing a numerical diagonalization of
$\mc{H}_{\trm{SOTI}}$, with the results shown in
Fig.~\ref{fig:surfaceStates_gapped_ens}(b).

In order to describe the low-energy physics of the $y=L_y$ surface, we
use the transformation $y \rightarrow L_y - y$,
$\vec{e}_j \rightarrow -\vec{e}_j$, which preserves the orientation of
the lateral surface. Under this transformation, the direction of the
Zeeman field changes sign such that the new low-energy Hamiltonian
becomes
\begin{align}
  \mc{H}^{xz}_{\trm{SOTI}}
  (y = L_y)
  & \approx \alpha A k_x \rho_z + t_1 B a k_z \rho_x
    \notag \\
  & -
    t_{\trm{Z}} C
    \left( n_y \rho_y + n_z \rho_z \right) .
\end{align}

Finally, the effective description of the two $yz$ surfaces is
 obtained by using the rotational symmetry in the $xy$ plane. We
  obtain
  \begin{align}
    \mc{H}^{yz}_{\trm{SOTI}}
    (x = 0)
    & \approx-\alpha A k_y \rho_z + t_1 B a k_z \rho_x
      \notag \\
    & +
      t_{\trm{Z}} C
      \left( n_x \rho_y + n_z \rho_z \right) ,
  \end{align}
  and
  \begin{align}
    \mc{H}^{yz}_{\trm{SOTI}}
    (x = L_x)
    & \approx-\alpha A k_y \rho_z + t_1 B a k_z \rho_x
      \notag \\
    & -
      t_{\trm{Z}} C
      \left( n_x \rho_y + n_z \rho_z \right) .
  \end{align}
  We note that the roles of the Zeeman field components along the
  $x$ and $y$ axis are interchanged. However, the mass term still gaps
  out the surface states and the sign of the mass term is opposite for
  the two surfaces.

\subsection{Emergence of the hinge states}

As we have seen previously, the effective low-energy description of
the surface states $\Ket{\Phi^{s}_{\pm}}$ of the 3DTI at zero Zeeman
field is characterized by the Dirac cone dispersion relation. The
staggered Zeeman field opens the gap at the Dirac point by inducing a
mass term with the amplitude $M_s$. The sign of $M_s$ is equal to the
projection of the Zeeman field vector $\vec{n}$ onto the
outward-pointing normal unit vector $\vec{v}^{s}_{\perp}$, which
changes sign between opposite surfaces. Hence, if all the surfaces of
the system are gapped, a set of gapless chiral modes has to emerge at
the hinge interfaces that connect the surfaces with opposite signs of
$M_s$. Such gapless chiral modes are denoted as hinge states. We use the
existence of the hinge states as a criterion to identify the emergence
of a SOTI topological phase.

An exact position of the hinge states is determined by calculating
  $\vec{v}^{s}_{\perp} \cdot \vec{n}$ on every surface. The presence
  of the hinge states is confirmed numerically by diagonalizing
  $\mc{H}_{\trm{SOTI}}$ in a geometry with OBCs along all three
  directions shown in Figs.~\ref{fig:surfaceStates_gapped_ens}(c)-(d)
  where we consider two most general orientations of the Zeeman
  field. In Fig.~\ref{fig:surfaceStates_gapped_ens}(c) the Zeeman
  field is taken to be parallel to the $xy$ surface, but making a
  finite angle with both $xz$ and $yz$ surfaces. As a result, the two
  hinge states emerge at the interfaces $(x=L_x, y=0)$ and
  $(x = 0, y = L_y)$ which connect the $xz$ and $yz$ surfaces with
  opposite signs of $\vec{v}^{s}_{\perp} \cdot \vec{n}$. The mass term
  on the $xy$ surfaces is zero, such that the whole surfaces remain
  gapless. In Fig.~\ref{fig:surfaceStates_gapped_ens}(d) the Zeeman
  field makes a finite angle with all three surfaces, gaping the $xy$
  surface out and leading to the emergence of four additional hinge
  states at $(x=L_x, z=0)$, $(y=L_y, z=0)$, $(x=0, z=L_z)$, and
  $(y=0, z=L_z)$. In both cases the gapless chiral state splits the
  whole system surface into two regions with opposite values of the
  mass $M_s$. Deforming the system geometry affects the exact position
  of the hinge state, but does not remove it. In particular, the
  rectangular geometry considered above can be deformed into a
  sphere. Then, the Zeeman field will split the sphere into two
  symmetric semi-spheres with opposite signs of $M_s$, leading to the
  emergence of a gapless chiral mode at the equator.

\subsection{Phase diagram}

\begin{figure}
  \centering \includegraphics[width=.99\columnwidth]
  {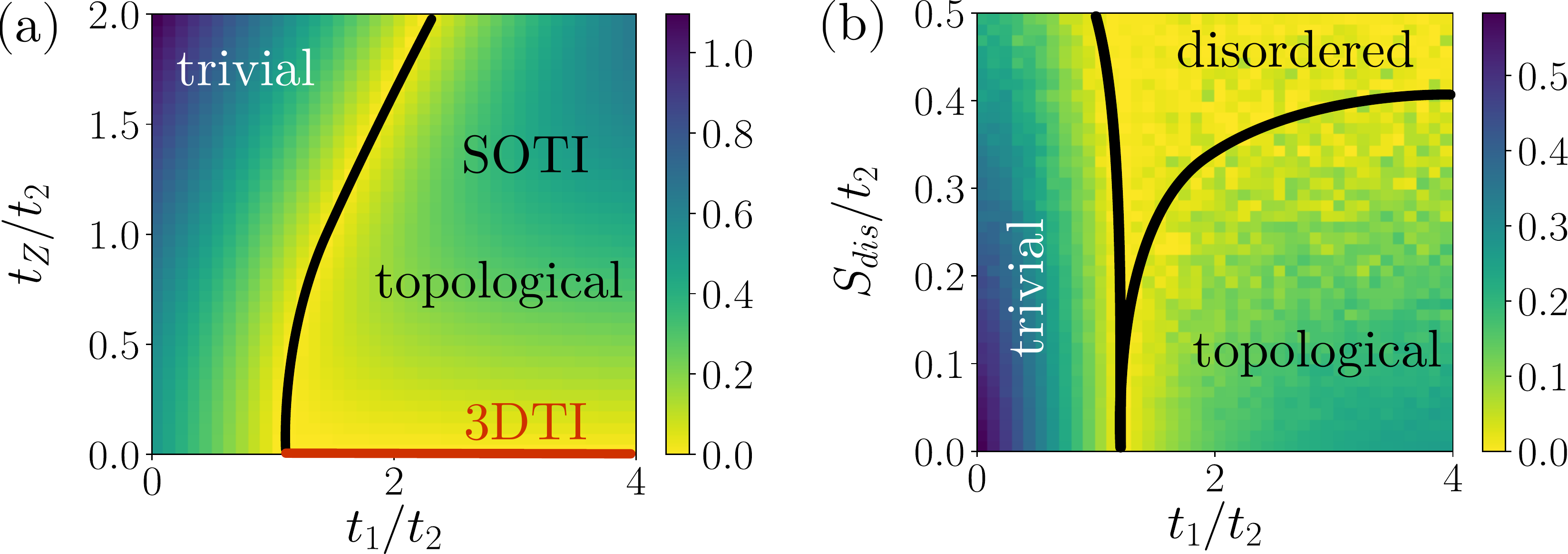}
  \caption{(a) Topological phase diagram showing boundaries between different phases as well as the size of the bulk gap (gap to the first excited
    state) as a function of the ratios $t_1 / t_2$ and
    $t_\trm{Z} / t_2$ for $t_2 = 0.5 E_{\trm{so}}$ (color coded in
    units of $E_{\trm{so}}$). The 3DTI phase (represented by the red
    line) exists in the regime $t_1 > t_2$ in the absence of Zeeman field. The
    SOTI phase emerges from the 3DTI at finite values of
    $t_{\trm{Z}}$. A critical value of the ratio $t_1 / t_2$
    separating the SOTI topological phase and the
    trivial gapped phase (represented by the black line) increases
    with the strength of the Zeeman field. The position of this
    critical line is not affected by the direction of the Zeeman
    field. (b) Topological phase diagram showing the stability of the topological
    phase against external perturbation and disorder as a function of
    the ratios $t_1 / t_2$ and $S_{\trm{dis}} / t_2$, where
    $S_{\trm{dis}} \coloneqq S_{\mu} = S_{\trm{Z}} =
    S_{\alpha}$. Every point on the phase diagram corresponds to a
    unique realization of the disorder. We see that the topological
    phase is stable even for  perturbations of  strength bigger
    than the bulk gap. When the perturbations become stronger, the bulk gap
    closes. The parameters of the simulation are
    $t_2 = 0.5 E_{\trm{so}}$ and $t_{\trm{Z}} = 0.3 E_{\trm{so}}$.}
  \label{fig:phaseDiagram}
\end{figure}

Previously, we obtained a simple low-energy description of the model
using the projection onto the space of surface states
$\Ket{\Phi^{s}_{\pm}}$. This description is valid only as long as
$t_{\trm{Z}} \ll t_1, t_2$. A complete phase diagram, including the
regime of strong staggered Zeeman field, is obtained numerically by
calculating the gap to the first excited state as a function of the
ratios $t_1 / t_2$ and $t_\trm{Z} / t_2$. The result of the
calculation is shown in Fig.~\ref{fig:phaseDiagram}(a). There we see
that the 3DTI phase is stable in the regime $t_1 > t_2$ at strictly
zero $t_{\trm{Z}}$, characterized by a gapless surface. At finite
values of $t_{\trm{Z}}$, the SOTI phase emerges from the 3DTI phase,
leading to the apparition of a gap to the first excited surface
state. Both 3DTI and SOTI phases are gapped in the bulk. We also
clearly see the critical line which separates the topologically
trivial phase at strong $t_2$ and the topological phase at strong
$t_1$. The Zeeman field shifts the phase boundary, so that higher
values of the ratio $t_1 / t_2$ are required to reach the SOTI
phase. Along this critical line, the system is gapless in the bulk. The
numerical calculations are performed for several directions of the
Zeeman field vector $\vec{n}$, but we find that it hardly affects the
position of the critical line, despite the strong anisotropy present in the system.

Finally, we discuss the stability of our setup with respect to
external perturbations and disorder. We verify numerically that the
topological states are not affected by the local on-site disorder such
as a fluctuating chemical potential or a Zeeman field generated by
impurities. In our calculations the disorder is implemented using a
uniform distribution with zero mean and standard deviations $S_{\mu}$
and $S_{\trm{Z}}$. Similarly, we check that the presence of
topological phases is not affected by the offset of fine tuned
parameters in different Rashba layers, such as the SOI amplitude
$\alpha$ and the chemical potential $\mu$. For this purpose, each
Rashba layer is equipped with a set of two random variables,
describing the offset of the SOI amplitude $\alpha$ and the chemical
potential $\mu$. The random variables are taken from a uniform
distributions with standard deviations $S_{\alpha}$ and
$S_{\mu}$. These random variables are constant within each Rashba
layer and only differ between different Rashba layers. The topological phase
diagram showing different phases as well as the values of bulk gaps (gap to the first excited state)
in a disordered system is presented in
Fig.~\ref{fig:phaseDiagram}(b). We find that the SOTI phase is stable
against the perturbations of a strength larger than the bulk gap.

\section{Modified setups \label{sec:surfaceField}}

In the last section, we consider two alternative approaches to
  generate the topological phase hosting hinge modes, which could facilitate the
  experimental realization.

\subsection{Magnetic proximity setup}

\begin{figure}
  \centering \includegraphics[width=.99\columnwidth]
  {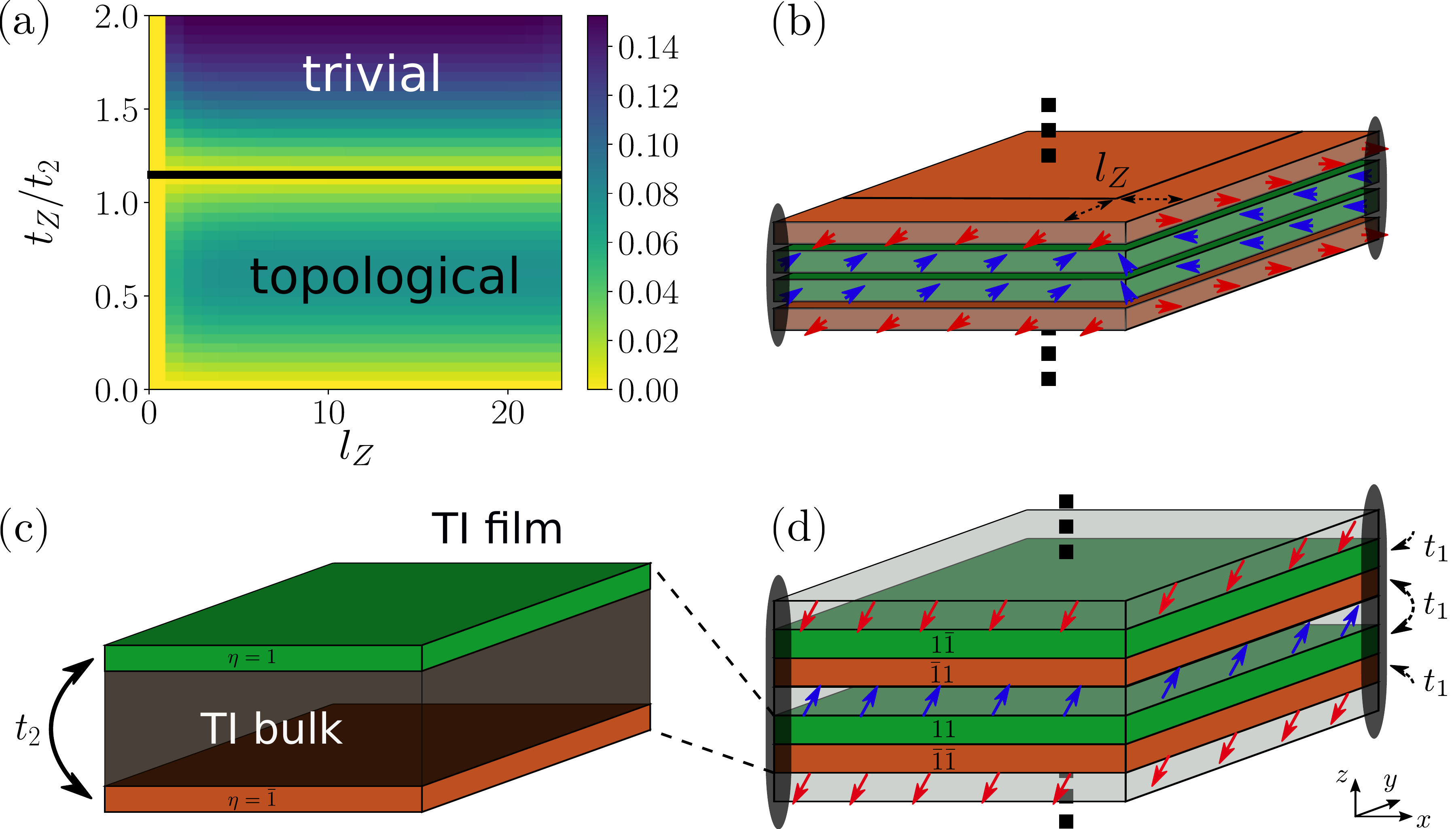}
  \caption{(a) Topological phase diagram showing boundaries between
    different phases as well as the size of the bulk gap as a function
    of the ratio $t_{\trm{Z}} / t_2$ and the penetration length of the
    Zeeman field $l_{\trm{Z}}$ for $t_1 = 1.5 t_2 = 0.75 E_{\trm{so}}$
    (color coded in units of $E_{\trm{so}}$). The
    surface states are completely gapped out even for a very small
    penetration length of the order of $2$ unit cells. We  note
    that the gap closes for $t_{\trm{Z}} \approx 1.1 t_2$
      (represented by the black line), indicating a topological phase
    transition. (b) Schematic representation of the Zeeman proximity
    setup where a staggered Zeeman field (represented by red and blue
    arrows) is generated by placing in proximity an antiferromagnetic
    material or by depositing magnetic adatoms on the surface of the
    Rashba layer heterostructure. (c) A scheme representing a thin TI
    film (only top and bottom gapless surface modes are
    shown). Because of the finite width of the film, gapless surface
    modes localized at the top and bottom surfaces are coupled via a
    $t_2$ term. (d) A schematics of a coupled-layer setup of thin TI
    films. Surface modes of the neighboring TIs with different
    helicities (represented by different colors) are coupled to each
    other via a $t_1$ term. Ferromagnetic layers are inserted between
    the TIs to generate a staggered Zeeman field. A pair of
      emerging hinge states is shown as two ellipsoids.}
  \label{fig:surfaceField}
\end{figure}

From the consideration above we know that the staggered Zeeman field has an effect only on the surface as the three-dimensional bulk modes are already gapped out by the tunneling between layers. Thus, we can consider a magnetic proximity setup where the
  staggered Zeeman field is induced only on the surface of the 3DTI
  and not in the bulk.
  Such a
  surface Zeeman field gaps out the states $\Ket{\Phi^{s}_{\pm}}$ by
  locally breaking the time-reversal symmetry at the surface. It leads
  to the emergence of hinge states without affecting the bulk
  properties of the system. The effect of the surface Zeeman field is
  tested numerically by calculating the gap to the first excited
  surface state, as a function of the ratio $t_{\trm{Z}} / t_2$ and
  the penetration length of the Zeeman field $l_{\trm{Z}}$. The result
  of the calculation is shown in Fig.~\ref{fig:surfaceField}(a). We
  find that even a very small penetration length, of the order of a
  few unit cells, is enough to fully gap out the surface
  states. Moreover, we observe that the topological phase transition
  occurs as a function of the surface Zeeman field. The position of
  the phase transition is not affected by the value of the penetration
  length $l_{\trm{Z}}$ and is independent of the bulk properties of
  the system.

An experimental realization of the staggered surface
Zeeman field can be achieved by using magnetic adatoms~\cite{Imps_LiuEtAl2009,
  Imps_ChenEtAl2010, Imps_BurkovBalents2011} or by placing an
antiferromagnetic material with a bicollinear antiferromagnetic
ordering close to the Rashba layer heterostructure, such that the size
of the magnetic unit cell along the $z$ direction of the ferromagnet
matches the distance $a$, as shown in Fig.~\ref{fig:surfaceField}(b).
The required antiferromagnetic materials have been studied both
theoretically~\cite{AFM_MaJiEtAl2009} and
experimentally~\cite{AFM_EnayatEtAl2014}. Such materials have also
been used experimentally in a proximity setup with a two-dimensional
TI~\cite{AFM_MannaEtAl2017}. This setup adds flexibility in
controlling the value of the mass term $M_s$ on different surfaces,
which can be achieved by adjusting the position of the
antiferromagnetic material or controlling the deposition of magnetic
atoms. This allows one to shift the position of the hinge state or even
generate several interfaces of zero $M_s$ on a single 3DTI surface.

\subsection{Coupled TI films}

Second, we propose an alternative coupled-layer construction
  where the 2DEGs with opposite dispersions are substituted by thin
  three-dimensional TI films. A thin TI film can be interpreted as a
  bilayer material hosting a pair of gapless surface states with Dirac
  dispersion relation at opposite surfaces [see
  Fig.~\ref{fig:surfaceField}(c)]. The finite width of the TI induces
  a tunnel coupling between the surface states, mediated by the bulk
  modes of the TI. We identify such a coupling term with the tunneling
  term $\mc{H}_2$ in Eq.~\eqref{eq:totalHamiltonian}. Moreover,
  stacking several TI films along the $z$ direction leads to another
  tunneling coupling between the surface states localized in
  neighboring TIs, which we identify with the tunneling term
  $\mc{H}_1$ in Eq.~\eqref{eq:totalHamiltonian}.

Similarly to the coupled 2DEG layer system, the TI layer setup
  requires a staggered Zeeman field to gap out the lateral
  surfaces. The Zeeman field has to change sign from one surface of
  the TI film to another, and keep the same sign at the interface
  between different TIs. The simplest way to realize such a Zeeman
  field consists in using thin ferromagnetic layers contained between
  the TI films. Alternatively, the effective Zeeman field can be
  generated using magnetic impurities. The schematic representation of
  the TI layer setup subjected to a staggered Zeeman field is shown in
  Fig.~\ref{fig:surfaceField}(d).

\section{Conclusions \label{sec:conclusion}}

To summarize, in this work we considered a system of coupled Rashba
layers subjected to a staggered Zeeman field. Such a model can be
realized experimentally by using 2DEGs with opposite signs of the
$g$-factor, by depositing magnetic adatoms or by placing an
antiferromagnet with a bicollinear antiferromagnetic order close to
the system surface. A similar effective model can also be obtained by
considering an array of coupled thin TI layers. At zero Zeeman field,
the system experiences a topological phase transition from the trivial
insulator to the strong 3DTI which hosts gapless modes at its
surface. Focusing on the low-energy degrees of freedom associated with
the surface states, we calculated perturbatively the effect of the
Zeeman term. We found that it leads to the emergence of a mass term
for gapless surface excitations that gaps them out. The amplitude of
the mass term is determined by the direction of the Zeeman field and
inevitably changes sign at opposite surfaces of the system. It leads
to the emergence of a zero-mass hinge interfaces hosting chiral
gapless hinge states, characteristic for the SOTI phase. We confirmed
numerically the presence of the hinge state and calculated the
topological phase diagram of our model. We found that the SOTI phase
is stable up to relatively large values of the Zeeman field. The
topological phase diagram also remains unaffected by external
perturbations and disorder of strength larger than the bulk gap.

\section*{Acknowledgments}

We gratefully acknowledge many useful discussions with Silas Hoffman and Yanick
Volpez. This work was supported by the Swiss National Science
Foundation, NCCR QSIT, and the Georg H. Endress foundation. This
project received funding from the European Union's Horizon 2020
research and innovation program (ERC Starting Grant, grant agreement
No 757725).

%%%%%%%%%%%%%%%%%%%%%%%%
%%%   Bibliography   %%%
%%%%%%%%%%%%%%%%%%%%%%%%

\bibliographystyle{unsrt}

%%%%%%%%%%%%%%%%%%%%%%%%%%%%%%%%%
%%%   Supplemental Material   %%%
%%%%%%%%%%%%%%%%%%%%%%%%%%%%%%%%%

\clearpage
\widetext

%% Merge with supplemental materials
%% Prefix a "S" to everything and reset the counter
\setcounter{section}{0}
\setcounter{equation}{0}
\setcounter{figure}{0}
\setcounter{page}{1}
\makeatletter
\renewcommand{\thesection}{Appendix \arabic{section}}
\renewcommand{\theequation}{A\arabic{equation}}
\renewcommand{\thefigure}{A\arabic{figure}}
\titleformat{\section}[hang]{\large\bfseries}{\thesection.}{5pt}{}

\section{Effect of the uniform Zeeman
  field~\label{sec:app-UniformZeeman}}

In the main text, we found that the staggered Zeeman field, described
by the term
\begin{align}
  \mc{H}_{\trm{Z}} = 
  t_{\trm{Z}} \vec{n} \cdot \vec{\sigma}\, \eta_z,
\end{align}
opens a gap in the spectrum of gapless surface states of a 3DTI,
associated with the mass term with the amplitude $M_s$. The mass $M_s$
changes sign between opposite faces, leading to the emergence of
gapless hinge states. In this Appendix, we show that the uniform
magnetic field described by a Zeeman term
\begin{align}
  \mc{H}_{\trm{ZU}}
  = 
  t_{\trm{Z}} \vec{n} \cdot \vec{\sigma} 
\end{align}
fails to open a gap in the spectrum of gapless surface states, and
does not allow one to generate hinge states.

\subsection{Top and bottom surfaces}

\begin{figure*}
  \centering
  \includegraphics[width=0.99\columnwidth]
  {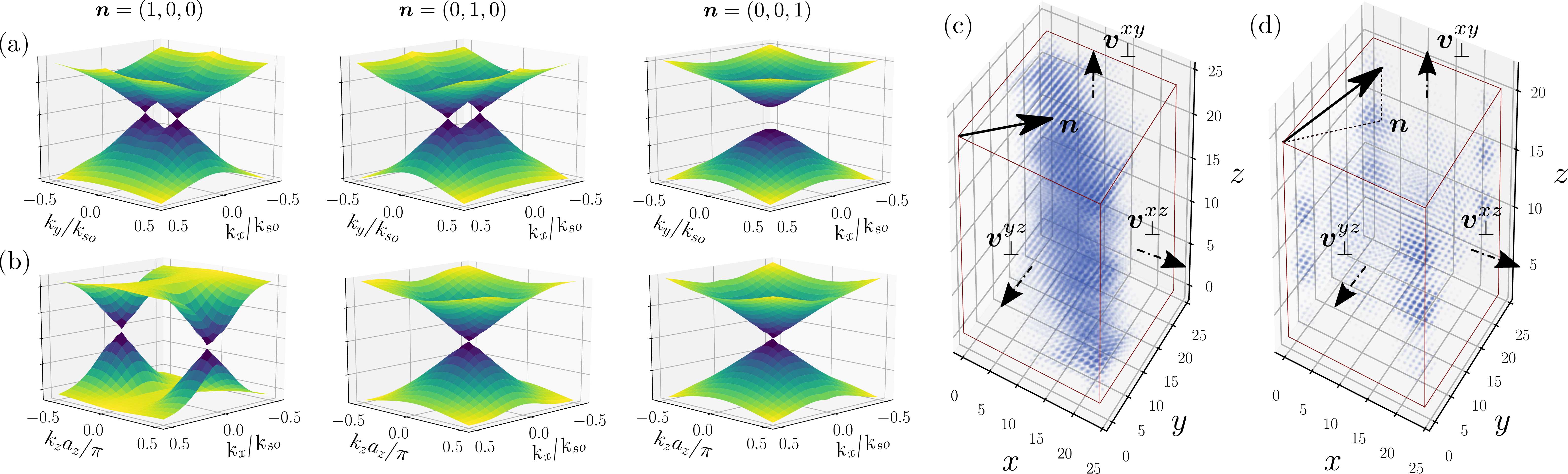}
  \caption{(a)-(b) The effect of the uniform Zeeman field on the
    dispersion relation of the surface states (a)
    $\Ket{\Phi^{xy}_{\pm}}$ and (b) $\Ket{\Phi^{xz}_{\pm}}$. The
    direction of the Zeeman field is indicated by the vector
    ${\vec{n}}$. The $z$ component of the Zeeman field gaps out the
    surface states at the top and bottom surfaces. The surface states
    at the lateral surfaces stay gapless for all the directions of the
    Zeeman field. Parameters of the simulations are
    $t_1 = 2 t_2 = E_{\trm{so}}$, $t_{\trm{Z}} = 0.3
    E_{\trm{so}}$. (c)-(d) Probability density of the state closest to
    the chemical potential, obtained in numerical simulations with
    OBCs along $x$, $y$, and $z$ directions. The black arrow indicates the
    direction of the Zeeman field. (c) If
    $\vec{n} = (1, 1, 0) / \sqrt{2}$, the Zeeman field fails to open
    a gap on any surface. (d) The Zeeman field with
    $\vec{n} = (1, 1, 1) / \sqrt{3}$ gaps out the $xy$ surfaces but
    leaves the lateral surfaces gapped. Parameters of the simulations
    are $t_1 = 2 t_2 = 4 E_{\trm{so}}$, and
    $t_{\trm{Z}} = 2 E_{\trm{so}}$.}
  \label{fig:surfaceStates_uniform_gapped_ens}
\end{figure*}

First, we focus on the top $xy$ surface and the surface states
$\Phi^{xy}_{\pm}(z)$ localized at $z=0$. We calculate the expectation
values of the Zeeman term $\mc{H}_{\trm{ZU}}$ in the neighborhood of
the Dirac point. We obtain the following diagonal contributions
\begin{align}
  \Braket{\Phi^{xy}_{\pm} |
  t_{\trm{Z}} n_x \sigma_x |
  \Phi^{xy}_{\pm}} =
  \Braket{\Phi^{xy}_{\pm} |
  t_{\trm{Z}} n_y \sigma_y |
  \Phi^{xy}_{\pm}} =
  0, \quad
  \Braket{\Phi^{xy}_{\pm} |
  t_{\trm{Z}} n_z \sigma_z |
  \Phi^{xy}_{\pm}} = \pm t_{\trm{Z}} n_z (t_1 - t_2) / \bar{t} .
\end{align}
Similarly, we calculate the off-diagonal contributions
\begin{align}
  &
    \Braket{\Phi^{xy}_{-} |
    t_{\trm{Z}} n_x \sigma_x |
    \Phi^{xy}_{+}} = -
    t_{\trm{Z}} n_x (t_1 - t_2) / \bar{t}
    , \quad
    \Braket{\Phi^{xy}_{-} |
    t_{\trm{Z}} n_y \sigma_y |
    \Phi^{xy}_{+}} = - i t_{\trm{Z}} n_y (t_1 - t_2) / \bar{t}
    , \quad
    \Braket{\Phi^{xy}_{-} |
    t_{\trm{Z}} n_z \sigma_z |
    \Phi^{xy}_{+}} = 0
    .
\end{align}
This allows us to write the effective low-energy Hamiltonian density
as follows
\begin{align}
  \mc{H}^{xy}_{\trm{SOTI}}
  (z = 0)
  \approx
  \alpha
  \left( \rho_x k_y - \rho_y k_x \right)
  (t_1 - t_2) / \bar{t}
  -
  t_{\trm{Z}}
  \left( n_x \rho_x + n_y \rho_y - n_z \rho_z \right)
  (t_1 - t_2) / \bar{t} .
\end{align}
We see that, similarly to the staggered Zeeman field, the $x$ and $y$
components of the uniform Zeeman field simply shift the position of
the Dirac cone in  momentum space, while the $z$ component gaps out
the surface states $\Ket{\Phi^{xy}_{\pm}}$. We note, however, that the
amplitude of the mass term generated by the uniform field is changed:
$M_{xy} = t_{\trm{Z}} n_z (t_1 - t_2) / \bar{t}$. Our analytical predictions are
confirmed by numerical calculations shown in
Fig.~\ref{fig:surfaceStates_uniform_gapped_ens}(a).

\subsection{Lateral surfaces}

Second, using the results of Sec.~\ref{sec:surfaceStatesLatZero},
we calculate the effect of the uniform Zeeman field on the surface
states localized on the lateral $xz$ surface at $y=0$. A numerical
calculation of the diagonal components of the expectation values
provides the following result
\begin{align}
  \Braket{\Phi^{xz}_{\pm} |
  t_{\trm{Z}} n_x \sigma_x |
  \Phi^{xz}_{\pm}} =
  \Braket{\Phi^{xz}_{\pm} |
  t_{\trm{Z}} n_y \sigma_y |
  \Phi^{xz}_{\pm}} =
  \Braket{\Phi^{xy}_{\pm} |
  t_{\trm{Z}} n_z \sigma_z |
  \Phi^{xy}_{\pm}} = 0.
\end{align}
Similarly, for the off-diagonal components we find
\begin{align}
  \Braket{\Phi^{xz}_{-} |
  t_{\trm{Z}} n_y \sigma_y |
  \Phi^{xz}_{+}} =
  \Braket{\Phi^{xz}_{-} |
  t_{\trm{Z}} n_z \sigma_z |
  \Phi^{xz}_{+}} =
  0, \quad
  \Braket{\Phi^{xy}_{-} |
  t_{\trm{Z}} n_x \sigma_x |
  \Phi^{xy}_{+}} = t_{\trm{Z}} n_x C.
\end{align}
This allows us to write down the effective low-energy Hamiltonian of
the $y=0$ surface
\begin{align}
  \mc{H}^{xz}_{\trm{SOTI}}
  (y = 0)
  \approx \alpha A k_x \rho_z + t_1 B a k_z \rho_x
  +
  t_{\trm{Z}} C n_x \rho_x .
\end{align}
Surprisingly, we find that the only component of the Zeeman field that
affects the low-energy degrees of freedom is the $x$ component. Its
main effect consists in shifting the Dirac cone along the $k_z$ axis
in  momentum space. We confirm these analytical predictions using
numerical simulations, shown in
Fig.~\ref{fig:surfaceStates_uniform_gapped_ens}(b). We see that the
gap remains closed for any direction of the Zeeman field. We also
note that the momentum shift originating from the uniform Zeeman
field is much larger than the one originating from the staggered
field.

\subsection{Numerical confirmation}

As a final confirmation of the effect of the uniform Zeeman field, we
perform numerical simulations of the system in a geometry with OBC
along all the three axis directions. The result of such simulations is
shown in Figs.~\ref{fig:surfaceStates_uniform_gapped_ens}(c)-(d). Similar
to the calculation in the main text, we consider two different
geometries, with the Zeeman field parallel to the $xy$ surface, and
with the Zeeman field making a finite angle with all the three
surfaces. As expected from our theoretical analysis, the $z$ component
of the uniform Zeeman field gaps out the surfaces states localized at
the top and bottom surfaces. No components of the Zeeman field open a
gap for the surfaces state localized at the lateral surfaces.

%%%%%%%%%%%%%%%%%%%% 
%%%     END      %%%
%%%%%%%%%%%%%%%%%%%%

\end{document}